\documentclass{ptephy_v1}

\usepackage{amsmath} 
\usepackage{amsthm} 

\theoremstyle{plain}
\newtheorem{theorem}{Theorem}

\newtheorem{definition}{Definition}

\begin{document}

\title{Attractive gravity probe surface with positive cosmological constant}



\author{Tetsuya Shiromizu${}^{1,2}$}
\author{Keisuke Izumi${}^{2,1}$}
\affil{${}^1$Department of Mathematics, Nagoya University, Nagoya 464-8602, Japan}
\affil{${}^2$Kobayashi-Maskawa Institute, Nagoya University, Nagoya 464-8602, Japan}


\begin{abstract}%
In four dimensional spacetimes with a positive cosmological constant, we introduce a new geometrical object 
associated with the cosmological horizon and then show the areal inequality. 
We also examine the attractive gravity probe surfaces as an indicator for the existence of attractive gravity. 
\end{abstract}

\subjectindex{xxxx, xxx}

\maketitle

\section{Introduction}

In spacetimes with a positive cosmological constant $\Lambda$, on a time-symmetric initial data, using the global monotonicity of 
a function associated with the quasi-local mass, the areal inequality for the area $A$ of the cosmological horizon, $A \leq 12\pi/\Lambda$, 
has been shown \cite{Boucher1984} (See also Ref. \cite{Maeda1998} for the argument based on the double null foliation and \cite{Gibbons1999} 
based on the second variation of the area) \footnote{We note that, in Refs. \cite{Boucher1984, Maeda1998}, certain specific foliation was taken. Similary, in 
Ref. \cite{Shiromizu1993} where discussed the area bound for the black hole in asymptotically  deSitter spacetimes, a specific foliation was adopted.}. 
Meanwhile, the loosely trapped surface (LTS)/attractive gravity probe surface (AGPS) was recently proposed as an indicator of 
strong gravity/attractive gravity \cite{Shiromizu2017, Izumi2021, Izumi2023}. The definition and existence of the AGPS are merely related to the 
positivity of the Arnowitt-Deser Misner (ADM) mass for asymptotically flat spacetimes 
(See Sec.~\ref{basics} and Appendix \ref{App}). In this paper, then, we adopt 
the essence of them to define a new geometrical object associated with the cosmological horizon, called cosmological gravity probe surface (CGPS) 
in spacetimes with positive cosmological constant and then we show the areal inequality. 
We see that the positivity of a quasi-local mass is essential for the areal inequality.  In addition, we introduce the 
AGPS for spacetimes with a positive cosmological constant and discuss the areal inequality. 
The AGPS can describe the black hole horizon as well as cosmological one. 
Indeed, one can see that the AGPS corresponds to them in certain limits of the parameter appearing in the definition of the AGPS. 
Since the definition of the CGPS/AGPS depends on the time slices, we shall introduce the two types. 

The rest of this paper is organized as follows. In Sec.~\ref{basics}, we will examine the Schwarzschild-deSitter spacetime as a typical example. 
In Sec.~\ref{SecCGPS} we will present the definition of the CGPSs for spacetimes with positive cosmological constant, and show the positivity of 
a quasi-local mass and the areal inequality. In Sec.~\ref{SecAGPS}, we will generalize the discussion to the AGPSs. 
Section~\ref{summary and discussion} is devoted to a summary and discussion. 
In the Appendix \ref{App}, we present the definition of the original 
AGPS for asymptotically flat spaces and discuss the relation to the positivity of the ADM mass.

\section{Lesson from Schwarzschild-deSitter spacetime}
\label{basics}

In this section, we analyze the Schwarzschild-deSitter spacetime as the simplest example 
and explore geometrical properties that may be satisfied in general cases. There are the two typical time slices, namely, 
static one and constant mean curvature one. 
Since we know that the behavior of geometrical quantities depends on the time slices, we will explore the 
Schwarzschild-deSitter spacetime in the two ways. 

\subsection{Static slice}

Let us start from the static slice of the Schwarzschild-deSitter spacetime  
\begin{equation}
ds^2=-f(r)dt^2+f^{-1}(r)dr^2+r^2d\Omega_2^2,
\end{equation}
where $f(r)=1-2m/r-(\Lambda /3) r^2$, $m$ is the mass parameter, $\Lambda$ is the positive cosmological constant and 
$d\Omega_2^2$ is the metric of the unit round 2-sphere. We focus on the cases with the horizons, where $f(r)=0$ has 
the root. 
Since the extrinsic curvature of the static slice vanishes, and thus the analysis 
here suggests that of the maximal slice discussed in subsections \ref{SecMS} and \ref{AGPSmax}.

As the loosely trapped surface \cite{Shiromizu2017} or attractive gravity probe surface \cite{Izumi2021, Izumi2023}, 
one examines the profile of the trace of the extrinsic curvature $k=(2/r){\sqrt {f(r)}}$ of 
$r=$ constant surfaces on the $t=$ constant hypersurface to specify the strength of 
gravity. This is characterized by the radial derivative of $k$
\begin{equation}
r^aD_a k=\frac{1}{r}\left(f'-\frac{2}{r}f \right)=-\frac{2}{r^2}\left(1-\frac{3m}{r} \right),
\end{equation}
where $r^a=f^{1/2}(\partial_r)^a$ and $D_a$ is the covariant derivative of the $t=$ constant hypersurfaces. 
Note that the cosmological constant does not appear in the above expression and it is easy to see that the maximum of $k$ 
is attained at the locus ($r=3m$) of the unstable circular orbit of photon
\footnote{This is the same with that in the Schwarzschild spacetime.}. 

Now we focus on the cosmological horizon at $r=r_C$. Using $f(r_C)=0$, one has 
\begin{equation}
r^aD_a k|_{r=r_C}=\frac{2m}{r^3_C}-\frac{2}{3}\Lambda. 
\end{equation}
Assuming the positivity of the mass $m$, one can see 
\begin{equation}
r^aD_a k|_{r=r_C} \geq -\frac{2}{3}\Lambda. \label{rDkrC}
\end{equation}
This geometric inequality may generally imply the positivity of mass 
(excluding the cosmological constant) within the cosmological horizon, and thus motivates us to introduce a 
new geometrical object associated with the cosmological horizon. 

Let us analyze the dependence of the mass in detail. One can also have in general 
\begin{equation}
r^aD_a k+\frac{k^2}{2}=\frac{2m}{r^3}-\frac{2}{3}\Lambda. \label{rDk+k^2/2=}
\end{equation}
Then, the positivity of the mass $m$ implies us 
\begin{equation}
r^aD_a k+\frac{k^2}{2} \geq -\frac{2}{3}\Lambda. \label{SSrDk+k^2/2}
\end{equation}
Or, it may be better to rewrite Eq.~\eqref{rDk+k^2/2=} in 
\begin{equation}
\frac{r^aD_a k}{{k^2} +\frac{4}{3}\Lambda} = -\frac12 + \frac1{2(r/m -2)}.
\label{dSSAGPS}
\end{equation}
Since the second term $[2(r/m -2)]^{-1}$ in the right-hand side is a decreasing function of the mass-normalized radius $r/m$, 
the ratio in the left-hand side is expected to indicate the mass-normalized distance from the gravitational source.
This property is the same as that of the original AGPS~\cite{Izumi2021,Izumi2023}, and 
gives a key for the definition of the AGPS for spacetimes with a positive cosmological constant. 
Note that on the cosmological horizon, since the areal radius $r_C$ is finite,
the right-hand side of Eq.~\eqref{dSSAGPS} is strictly larger than $-1/2$. 
The difference from $-1/2$ stems from the second term in the right-hand side, which is a function of the mass-normalized distance. 
This implies that the gravitational attraction affects the geometry of the cosmological horizon, 
as opposed to the geometry of the spatial infinity in asymptotically flat (or anti-deSitter) spaces.

\subsection{Constant mean curvature slice}

The accelerated expansion of the universe is approximated by the flat chart of the deSitter spacetime, 
where the extrinsic curvature  $K_{ab}$ of the time-constant hypersurface becomes $K_{ab}=H q_{ab}$. 
Here, $q_{ab}$ is the induced metric of the hypersurface and $H:=\sqrt{\Lambda/3}$. 
The Schwarzschild-deSitter allows us to simply incorporate the effects of mass or energy, and the slice 
corresponding to that in the flat chart of the deSitter spacetime is the constant mean curvature slice of the Schwarzschild-deSitter spacetime,   
whose metric is rewritten as \footnote{The relation between 
the current coordinates ($\tau$, $\rho$) and the ($t, r$) is given by 
$r=a \rho \left(1+\frac{m}{2a\rho} \right)^2$ and $t+h(r)=\tau$, where $h'(r)=Hr^2/(f(r){\sqrt {r(r-2m)}})$. }\cite{McVittie1933}
\begin{equation}
ds^2=-\left( \frac{1-\frac{m}{2a\rho}}{1+\frac{m}{2a\rho}} \right)^2d\tau^2+\left( 1+\frac{m}{2a\rho} \right)^4 a^2(d\rho^2+\rho^2d\Omega_2^2),
\end{equation}
where $a=e^{H \tau}$. 
The analysis here corresponds to that of a constant mean curvature slice 
discussed in subsections \ref{SecCMCS} and \ref{AGPScmc}.

As the previous subsection, we compute the trace of the extrinsic curvature and 
its derivative of the $\rho=$ constant surface on the $\tau=$ constant hypersurface as 
\begin{equation}
k=\frac{2}{a\rho}\frac{1-\frac{m}{2a\rho}}{\left( 1+\frac{m}{2a\rho} \right)^3}
\end{equation}
and
\begin{equation}
r^a D_ak=-\frac{2}{(a\rho)^2}\frac{1-\frac{2 m}{a\rho}+\frac{1}{4}\left(\frac{m}{a\rho} \right)^2}{\left( 1+\frac{m}{2a\rho} \right)^6},
\end{equation}
respectively. In the above, $D_a$ is the covariant derivative of the $\tau=$ constant hypersurfaces. 
Moreover, we have 
\begin{equation}
r^a D_ak+\frac{1}{2}k^2=2\frac{m}{(a\rho)^3}\frac{1}{\left( 1+\frac{m}{2a\rho} \right)^6}. \label{rDk+k^2/2}
\end{equation}
It is easy to see that the extrinsic curvature $K_{ab}$ of the $\tau=$ constant hypersurface with the induced metric $q_{ab}$ is 
\begin{equation}
K_{ab}=Hq_{ab}
\end{equation}
and then $K:=K^a_a=3H$. 
Here note that the geometrical object associated with the cosmological horizon is specified by the vanishing of the 
expansion rate $\theta_-$ for ingoing null geodesics and then 
\begin{equation}
k=2H \label{k=2H}
\end{equation}
holds on the cosmological horizon because of 
\begin{equation}
\theta_- \propto K_{ab}(q^{ab}-r^ar^b)-k. \label{expansion-}
\end{equation}
Then, together with the condition Eq. (\ref{k=2H}) and the positivity of the mass $m$,  
Eq. (\ref{rDk+k^2/2}) implies us 
\begin{equation}
r^a D_ak+\frac{2}{3}\Lambda \geq 0. 
\end{equation}
This gives us a key for the definition of the surface associated with the cosmological horizon. 

In general, the positivity of the mass $m$ gives us 
\begin{equation}
r^a D_ak+\frac{1}{2}k^2 \geq 0.  \label{rDk+k^2/22}
\end{equation}
The ratio between $r^a D_ak$ and $k^2$ is 
\begin{equation}
\frac{r^a D_ak}{k^2} = -\frac12 + \frac{a\rho/m}{2(a\rho/m-1/2)^2}.
\label{CFSratio}
\end{equation}
Since the second term in the right-hand side is a decreasing function of the mass-normalized radius $a\rho/m$, 
as in the case of asymptotically flat spacetime, the AGPS for asymptotically deSitter spacetimes can be introduced. 
The mass-normalized radius $a\rho/m$ has a finite value on the cosmological horizon,  
the second term in the right-hand of Eq.~\eqref{CFSratio} is positive. 
This implies that the cosmological horizon is affected by the existence of mass, {\it i.e.}, 
the gravitational attractive force.

\section{Cosmological gravity probe surfaces}
\label{SecCGPS}

In this section, bearing the consideration on the Schwarzschild-deSitter spacetime in mind, 
we introduce the two types of the cosmological gravity probe surface (CGPS) as the indicator of the existence of the 
cosmological horizon in spacetimes with positive cosmological constant and show the areal inequality. 
On the way, we see that the condition in the definition of the CGPS is related to the positivity of the mass. 

Hereafter we denote the trace of the extrinsic curvature of two-surfaces in spacelike hypersurfaces, the outward unit normal vector 
to two-surfaces and the covariant derivative with respect to spacelike hypersurfaces by $k$, $r^a$ and $D_a$, respectively.

\subsection{Cosmological gravity probe surfaces in maximal slice}
\label{SecMS}


Inspired by the observation for the Schwarzschild-deSitter spacetime in the static slice, 
we propose {\it the cosmological gravity prove surface in a maximal spacelike hypersurface (CGPS-max)}.  
\begin{definition}
Let us consider the four-dimensional spacetime following the Einstein equation with the positive cosmological constant $\Lambda$. Then 
we define the {\it cosmological gravity probe surface in a maximal spacelike hypersurface } $\Sigma$ 
as a compact surface $S_C$ satisfying $k|_{S_C}=0$ and 
\begin{equation}
r^aD_a k|_{S_C} \geq -\frac{2}{3}\Lambda. \label{CGPS}
\end{equation}
\end{definition}
Here note that the definition depends on the foliation/lapse near the surface. 
However, we do not specify a foliation, 
because the following discussion is valid as long as 
there is a surface that satisfies the conditions for a foliation.
This is common in all definitions appeared later. 

On the maximal hypersurface, the Hamiltonian constraint shows us 
\begin{equation}
{}^{(3)}R=K_{ab}K^{ab}+16 \pi T_{ab}n^an^b+2\Lambda,
\end{equation}
where $K_{ab}$ is the extrinsic curvature of $\Sigma$, $T_{ab}$ is the energy-momentum tensor and $n^a$ is the unit normal vector to $\Sigma$. 
If the positivity of the energy density, $T_{ab}n^an^b \geq 0$, is assumed, it leads us the condition 
\begin{equation}
{}^{(3)}R\ge 2\Lambda. \label{3Ricci}
\end{equation}

Let us define the following quasi-local mass for a 2-surface $S$ in 
spacelike hypersurface~\cite{Boucher1984, Nakao1995, Maeda1998}
\begin{equation}
E_s(S)=\frac{A^{1/2}(S)}{64\pi^{3/2}}\int_{S} \left(2{}^{(2)}R-k^2-\frac{4}{3}\Lambda  \right)dA, \label{Es}
\end{equation}
where $A(S)$ is the area and ${}^{(2)}R$ is the Ricci scalar of $S$. For an $r=$ constant surface on the static slice of the Schwarzschild-deSitter spacetime, it gives us $m$. 
For $\Lambda=0$, it becomes the Geroch mass used to show the positivity of the Arnowitt-Deser-Misner(ADM) mass in 
asymptotically flat spacetimes \cite{Geroch1973}. 

We give a theorem which shows that inequality~\eqref{SSrDk+k^2/2} guarantees the nonnegativity of the quasi-local mass $E_s$ in any space satisfying Eq.~\eqref{3Ricci}. 
\begin{theorem}
Let us consider the four-dimensional spacetime following the Einstein equation with the positive cosmological constant $\Lambda$ and 
satisfying the dominant energy condition. Let $\Sigma$ be a maximal spacelike hypersurface. 
Then, a surface $S_0$ in $\Sigma$ satisfying \footnote{
In equation (\ref{CFSratio}), for regions where $r$ is sufficiently small, i.e., near the gravity source, 
the right-hand side can be much larger than -1/2. This makes equation (\ref{gCGPS}) easier to satisfy. 
Therefore, with a certain amount of  matter fields, we can expect to obtain a surface that satisfies Eq. (\ref{gCGPS}). 
Moreover, it is easy to see that the theorem holds with condition (\ref{gCGPS}) relaxed to the surface integral form, that is,
$\int_{S_0}r^aD_a kdA \geq -\int_{S_0} (k^2/2 +2\Lambda/3 )dA$.
}
\begin{equation}
r^aD_a k|_{S_0} \geq -\frac12 k^2 -\frac{2}{3}\Lambda, \label{gCGPS}
\end{equation}
has topology $S^2$ and the quasi-local mass $E_s(S_0)$ is nonnegative. 
And, $E_s(S_0)=0$ holds if and only if the lapse function for the outward normal direction to $S_0$ 
is constant on $S_0$, $S_0$ is totally umbilic, 
and the equalities in Eqs.~\eqref{3Ricci} and \eqref{gCGPS} hold on $S_0$. 
\end{theorem}

The proof is rather simple. The key equation is the following identity
\begin{equation}
r^aD_ak=-\varphi^{-1}{\cal D}^2 \varphi- \frac12 {}^{(3)}R+\frac{1}{2}{}^{(2)}R-\frac{3}{4}k^2 -\frac12 \tilde k_{ab} \tilde k^{ab}, \label{rDk}
\end{equation}
where ${\cal D}_a$ and $\tilde k_{ab}$ are the covariant derivative and the traceless part of the extrinsic curvature $k_{ab}$ of $S_0$ in $\Sigma$, respectively. 
$\varphi$ is the lapse function for the outward normal direction to $S_0$ in $\Sigma$. 
Taking the surface integration over $S_0$ for Eq. (\ref{rDk}), the condition (\ref{gCGPS}) implies  
\begin{equation}
\int_{S_0} \left( \frac{1}{2}{}^{(3)}R-\frac{2}{3}\Lambda+\frac14 k^2+\frac{1}{2}\tilde k_{ab} \tilde k^{ab}+\varphi^{-2}({\cal D} \varphi)^2   \right) dA \leq 
\frac{1}{2} \int_{S_0}{}^{(2)}RdA.
\end{equation}
Then, Eq. (\ref{3Ricci}) shows us 
\begin{equation}
 0 < \int_{S_0} \left( \frac{1}{3}\Lambda+\frac14 k^2+\frac{1}{2}\tilde k_{ab} \tilde k^{ab}+\varphi^{-2}({\cal D} \varphi)^2   \right) dA \leq 
\frac{1}{2} \int_{S_0}{}^{(2)}RdA.
\label{intCGPS}
\end{equation}
Thus, the Gauss-Bonnet theorem tells us that the topology of $S_0$ is $S^2$, and then it gives 
\begin{equation}
\int_{S_0}{}^{(2)}RdA=8\pi. 
\end{equation}
Moreover, Eq.~\eqref{intCGPS} is rewritten as 
\begin{equation}
0 \le \int_{S_0} \left( \frac{1}{2}\tilde k_{ab} \tilde k^{ab}+\varphi^{-2}({\cal D} \varphi)^2   \right) dA \leq  
\int_{S_0} \left( \frac{1}{2} {}^{(2)}R -\frac14 k^2 - \frac{1}{3}\Lambda \right)  dA = \frac{16 \pi^{3/2}}{A^{1/2}(S_0)} E_s(S_0).
\end{equation}
This shows the nonnegativity of $E(S_0)$,
corresponding to imposing the positivity of the mass for the Schwarzschild-deSitter spacetime as discussed in Sec. 2. 1. 
The equalities hold if and only if the equalities in Eqs.~\eqref{gCGPS} and \eqref{3Ricci} hold, ${\cal D}_a \varphi=0$, 
that is, the lapse function is constant on $S_0$, and $\tilde k_{ab}=0$, that is, $S_0$ is totally umbilic. 
Note that, the equality in Eq.~\eqref{3Ricci} holds if and only if 
$K_{ab}=0$ and $T_{ab}n^a n^b=0$ are satisfied

Now, we show an upper bound for the area of $S_0$ defined in the Theorem 1,
\begin{theorem}
Let us consider the four-dimensional spacetime following the Einstein equation with the positive cosmological constant $\Lambda$ and 
satisfying the dominant energy condition. 
Let $\Sigma$ be a maximal spacelike hypersurface. 
Then, the area $A(S_0)$ of a surface $S_0$ with Eq.~\eqref{gCGPS} in $\Sigma$ satisfies
\begin{equation}
A(S_0) \leq \frac{12\pi}{\Lambda}. \label{arealineq}
\end{equation}
And, the equality holds if and only if $S_0$ is a CGPS-max and $E_s(S_0)=0$ holds.
\end{theorem}

The statement follows from  only the facts that $E_s(S_0)$ is nonnegative and the topology of $S_0$ 
is $S^2$~\cite{Boucher1984}\footnote{The fact that the $S_0$ has $S^2$ topology can be shown from 
the nonnegativity of $E_s(S_0)$.}. 
The nonnegativity of $E_s(S_0)$ gives 
\begin{equation}
(0 \leq) \frac{4}{3}\Lambda A(S_0) 
=  \int_{S_0} \left( \frac{4}{3}\Lambda \right)  dA 
 \leq  \int_{S_0} \left(  k^2 + \frac{4}{3}\Lambda \right)  dA 
 \leq  \int_{S_0} 2 {}^{(2)}R  dA 
 =16 \pi
 \end{equation}
This gives Eq.~\eqref{arealineq}. 
Equalities hold if and only if $k=0$ and $E_s(S_0)=0$ hold.
Note that the inequality~\eqref{arealineq} holds for the CGPS-max.

\subsection{Cosmological gravity probe surfaces in constant mean curvature slice}
\label{SecCMCS}


In this subsection, bearing the observation for the Schwarzschild-deSitter spacetime in constant mean curvature slice, 
we propose another type of the CGPS, that is, {\it the CGPS in constant mean curvature slice (CGPS-cmc)}. 
\begin{definition}
Let us consider the four-dimensional spacetime following the Einstein equation with the positive cosmological constant $\Lambda$. 
Then we define the cosmological gravity probe surface in spacelike hypersurface with the trace of 
the extrinsic curvature $K=3{\sqrt {\Lambda/ 3}}$ as a compact surface $S_c$ satisfying $k|_{S_c}=2{\sqrt {\Lambda/ 3}}$ and 
\footnote{As Theorem 1, it may be better to relaxed to the surface integral form, that is, $ \int_{S_c}r^aD_a k dA \geq -\int_{S_c}(k^2/2)dA
=-(2\Lambda/3)A_c$. In a similar way, the conditions appeared in almost of all definitions/theorems hereafter can be replaced by the surface 
integral form.} 
\begin{equation}
r^aD_a k|_{S_c} \geq -\frac{k^2|_{S_c}}{2}=-\frac{2}{3}\Lambda. \label{CGPS2}
\end{equation}
\end{definition}

On the current constant mean curvature surface, the Hamiltonian constraint shows us 
\begin{equation}
{}^{(3)}R=\tilde K_{ab} \tilde K^{ab}+16 \pi T_{ab}n^an^b,
\end{equation}
where $\tilde K_{ab}$ is the traceless part of the extrinsic curvature of $\Sigma$. 
Under the assumption of the positivity of the energy density, $T_{ab}n^an^b \geq 0$, it leads us the condition 
\begin{equation}
{}^{(3)}R\ge 0. \label{3Riccige0}
\end{equation}

Since the CGPS-cmc is inspired by the observation for the Schwarzschild-deSitter spacetime in the constant mean curvature slices, 
which are asymptotically flat, 
one may employ the Geroch mass \cite{Geroch1973}
\begin{equation}
E_f(S)=\frac{A^{1/2}(S)}{64\pi^{3/2}}\int_{S} \left(2{}^{(2)}R-k^2  \right)dA. \label{Ef}
\end{equation}
For an $r=$ constant surface in the constant mean curvature slice of the Schwarzschild-deSitter spacetime, it gives us $m$. If we consider the spacetime following the Einstein equation with 
the positive cosmological constant and satisfying the dominant energy condition, 
the geometrical identity~\eqref{rDk}, the condition~\eqref{CGPS2} and Eq. (\ref{3Riccige0}) give
\begin{equation}
0< \frac12 \int_{S_c}  k^2 dA \leq  
\int_{S_c} \left( {}^{(3)}R+\frac12 k^2+\tilde k_{ab} \tilde k^{ab}+2\varphi^{-2}({\cal D} \varphi)^2   \right) dA \leq 
\int_{S_c}{}^{(2)}RdA = 8\pi,
\label{intCGPScmcf}
\end{equation}
where in the last equality we use the Gauss-Bonnet theorem, that is, it fixes the topology of $S_c$ to be $S^2$ because of 
the positivity of $\int_{S_c}{}^{(2)}RdA$ and then the integration becomes $8\pi$.
On a CGPS-cmc, since the mean curvature of $S_c$ is constant by definition,  the Geroch mass is written as
\begin{equation}
E_f(S_c)=\frac{A^{1/2}(S_c)}{64\pi^{3/2}} \left( 16\pi -\frac{4}{3}\Lambda A(S_c) \right).
\label{EfSc}
\end{equation}
Moreover, Eq.~\eqref{intCGPScmcf} shows the nonnegativity of the Geroch mass $E_f(S_c)\ge0$. 
Note that, to show the nonnegativity of the Geroch mass, it is easy to see that only the inequality in Eq.~\eqref{CGPS2} without 
specifying the value of $k$, that is 
\begin{equation}
r^aD_a k|_{S_c} \geq -\frac{k^2}{2}
\end{equation}
is relevant~\cite{Izumi2021,Izumi2023}.
Therefore, the condition \eqref{CGPS2} guarantees the nonnegativity of the Geroch mass, corresponding to 
imposing the positivity of the mass for the Schwarzschild-deSitter spacetime as discussed in Sec. 2.2. 

The nonnegativity of the Geroch mass directly gives the areal inequality for the CGPS-cmc, {\it i.e.}, 
the nonnegativity of \eqref{EfSc} is equivalent to
\begin{equation}
A(S_c) \leq \frac{12\pi}{\Lambda}. \label{arealineq2}
\end{equation}

Now, we summarize the facts obtained in this subsection as the following theorem.
\begin{theorem}
Let us consider the four-dimensional spacetime following the Einstein equation with the positive cosmological constant $\Lambda$. 
Then, for the cosmological gravity probe surface $S_c$ in spacelike hypersurface with the trace of 
the extrinsic curvature $K=3{\sqrt {\Lambda/ 3}}$, the Geroch mass $E_f(S_c)$ of any CGPS-cmc $S_c$ is nonnegative and the area of $S_c$ satisfies Eq.~\eqref{arealineq2}. 
Equality holds if and only if the equality in Eq.~\eqref{CGPS2} holds, $^{(3)}R=0$ is satisfied on $S_c$, the 
lapse function for the outward normal direction to $S_c$ in $\Sigma$ is constant on $S_c$, and $S_c$ is totally 
umbilic in $\Sigma$.
\end{theorem}

\section{Attractive gravity probe surface in spacetimes with positive cosmological constant}
\label{SecAGPS}

The attractive gravity probe surface (AGPS) was introduced as an indicator for the presence of the 
attractive gravity in asymptotically flat spaces~\cite{Izumi2021,Izumi2023}. 
The strength of gravitational field or distance from gravitational sources are described by a parameter $\alpha$. 
The effect of the gravitational field is suppressed near the spatial infinity corresponding to $\alpha \to -1/2$, 
while at a finite distance from the gravitational sources ($\alpha > -1/2$) the gravitational field affects to 
the upper bound of the area of the AGPS. 

Let us generalize the notion of AGPS for spacetimes with a 
positive cosmological constant.
The maximal slice in spacetimes with a positive cosmological constant has a lower bound of the Ricci scalar $^{(3)}R\ge 2\Lambda$. 
This has a naive tendency which makes the volume of a space finite. 
Therefore, the distance from the gravitational source becomes finite and we expect to see the effect of 
the gravitational field, {\it i.e.}, the parameter $\alpha$. 
On the other hand, the size of cosmological horizon in the constant mean curvature slice is finite, 
and thus, the effect of the gravitational field becomes important there. 
Hence, the attractive gravity is expected to be probed.

In this section, we introduce the attractive gravity probe surface for spacetimes with a positive cosmological constant. 
We have the two subsections discussing general cases depending on the choice of the slice.

\subsection{Attractive gravity probe surface in maximal slice}
\label{AGPSmax}

First, we present the definition of {\it the AGPS in maximal spacelike hypersurface (AGPS$_\Lambda$-max)} as follows.
\begin{definition}
Let us consider the four-dimensional spacetime following the Einstein equation with the positive cosmological constant $\Lambda$. 
Then, we define the attractive gravity probe surface $S_\alpha$ in a maximal spacelike hypersurface $\Sigma$ as a compact surface satisfying $k|_{S_\alpha} \geq 0$ and 
\footnote{In our previous paper~\cite{Izumi2021} discussing cases with a negative cosmological constant, 
we define the AGPS with a parameter $\alpha$ as a surface satisfying not Eq.~\eqref{AGPSdS} but 
\begin{equation*}
r^aD_a k|_{S_\alpha} \geq \alpha k^2.
\end{equation*}
However, since the definition based on Eq.~\eqref{AGPSdS} gives a simple result, here we use it.}
\begin{equation}
r^aD_a k|_{S_\alpha} \geq \alpha \left(k^2+ \frac{4}{3}\Lambda \right), \label{AGPSdS}
\end{equation}
where $\alpha$ is a constant greater than $-1/2$.
\end{definition}

This definition is motivated by the consideration on the Schwarzschild-deSitter spacetime (for example, see Eq. (\ref{dSSAGPS})). 
If we set $\Lambda=0$, it coincides with the original one proposed for asymptotically flat spaces \cite{Izumi2021}. Now we present the 
following theorem.
\begin{theorem}
Let us consider the four-dimensional spacetime following the Einstein equation with the positive cosmological constant $\Lambda$ and 
satisfying the dominant energy condition. 
Then, an AGPS$_\Lambda$-max $S_\alpha$ with a parameter $\alpha$ in a maximal spacelike hypersurface $\Sigma$ has topology $S^2$ 
and the quasi-local mass $E_s(S_\alpha)$ satisfies
\begin{equation}
E_s(S_\alpha) \geq \frac{1+2\alpha}{3+4\alpha}\left(\frac{A_\alpha}{4\pi}\right)^{1/2} (\geq 0),
\label{BoundEa}
\end{equation}
where $A_\alpha$ is the area of $S_\alpha$.
Equality holds if and only if the lapse function for the outward normal direction to $S_\alpha$ 
is constant on $S_\alpha$, $S_\alpha$ is totally umbilic, $^{(3)}R=2\Lambda$ is satisfied, and the equality in Eq.~\eqref{AGPSdS} holds. 
\end{theorem}

The proof is parallel to that of Theorem 2. 
Instead of Eq.~\eqref{intCGPS}, we have
\begin{equation}
 0 < \int_{S_\alpha} \left[\left(\alpha + \frac34\right) \left(k^2 + \frac43\Lambda \right)+\frac{1}{2}\tilde k_{ab} \tilde k^{ab}+\varphi^{-2}({\cal D} \varphi)^2   \right] dA \leq 
\frac{1}{2} \int_{S_\alpha}{}^{(2)}RdA.
\label{intAGPS}
\end{equation}
The Gauss-Bonnet theorem tells us that the topology of $S_\alpha$ is $S^2$ and 
\begin{equation}
\int_{S_\alpha}{}^{(2)}RdA=8\pi. 
\end{equation}
Moreover, Eq.~\eqref{intAGPS} gives
\begin{equation}
0<\int_{S_\alpha}  \left(k^2 + \frac43\Lambda \right) dA \leq 
\frac{1}{2} \left(\alpha + \frac34\right)^{-1} \int_{S_\alpha}{}^{(2)}RdA = \frac{16}{3+4\alpha}\pi.
\label{intAGPS2}
\end{equation}
Then, the quasi-local mass $E(S_\alpha)$ is bounded below as
\begin{equation}
E_s(S_\alpha) = \frac{A_\alpha^{1/2}}{64 \pi^{3/2}}\int_{S_\alpha} \left(2 {}^{(2)}R - k^2 -\frac43\Lambda\right) dA \ge 
\frac{A_\alpha^{1/2}}{64 \pi^{3/2}} \left(16\pi - \frac{16}{3+4\alpha}\pi\right) 
=\frac{1+2\alpha}{3+4\alpha}\left(\frac{A_\alpha}{4\pi}\right)^{1/2}.
\end{equation}
Equality holds if and only if all inequalities in the proof become equalities, that is, $\tilde k_{ab}=0$, ${\cal D}_a \varphi=0$, $^{(3)}R=2\Lambda$ and 
equality holds in Eq.~\eqref{AGPSdS}. 
Note that $^{(3)}R=2\Lambda$ holds if and only if 
$K_{ab}=0$ and $T_{ab}n^a n^b=0$ are satisfied.

Inequality~\eqref{AGPSdS} can be expressed as
\begin{equation}
A_\alpha \leq 4\pi \left( \frac{3+4\alpha}{1+2\alpha}E_s(S_\alpha) \right)^2. \label{PIAGPS1}
\end{equation}
Except for the fact that $E_s(S_\alpha)$ is not mass in general \footnote{Since there is no asymptotic time translation symmetry in 
asymptotically deSitter spacetimes, it is difficult to define the mass, and the relation between the quasi-local 
mass $E_s(S_\alpha)$ and the observable mass is not clear.}, 
the apparent expression is the same with the inequality obtained for asymptotically flat spacetimes \cite{Izumi2021, Izumi2023}.

Next, we show the following upper bound for the area of the AGPS$_\Lambda$-max.
\begin{theorem}
Let us consider the four-dimensional spacetime following the Einstein equation with the positive cosmological constant $\Lambda$ and 
satisfying the dominant energy condition. 
Then, the area $A_\alpha$ of an AGPS$_\Lambda$-max satisfies
\begin{equation}
A_\alpha \leq \frac{12\pi}{3+4\alpha}\frac{1}{\Lambda}. \label{arealineq3}
\end{equation}
Equality holds if and only if $k|_{S_\alpha}=0$ and $E_s(S_\alpha)=0$ hold.
\end{theorem}

The proof is similar to that of Theorem 2. 
Inequality \eqref{AGPSdS} gives us
\begin{equation}
(0 \leq) \Lambda A_\alpha 
=  \int_{S_\alpha} \Lambda dA 
 \leq \frac34 \int_{S_\alpha} \left(  k^2 + \frac{4}{3}\Lambda \right)  dA 
 \leq \frac34 \int_{S_\alpha} 2 {}^{(2)}R  dA -24\pi \frac{1+2\alpha}{3+4\alpha}
 = \frac{12\pi}{3+4\alpha}.
 \end{equation}
Equalities in the first and second inequalities hold if and only if $k=0$ and $E_s(S_\alpha)=0$ hold, respectively.

Note that one can recover Eq. (\ref{arealineq}) in the limit of $\alpha \to -1/2$. Moreover, Eq.~(\ref{arealineq3}) also gives us 
\begin{equation}
A_\alpha \leq \frac{4\pi}{\Lambda} \label{arealineqBH}
\end{equation}
in the limit of $\alpha \to 0$, 
that is, $S_\alpha$ corresponds to the loosely trapped surface (LTS) defined in Ref. \cite{Shiromizu2017}. 
The upper bound for the LTS in asymptotically deSitter spacetimes is natural to be the same with that for a black hole horizon shown 
in Ref. \cite{Shiromizu1993, Hayward1994, Maeda1998, Gibbons1999, Galloway2018, Shiromizu2022}.  

Note that, the condition~\eqref{AGPSdS} is stronger than Eq.~\eqref{gCGPS}, 
which is the reason why we have the stronger results than those in Sec.~\ref{SecMS}.
As we have seen in Eq.~\eqref{dSSAGPS}, $r^aD_ak/(k^2+4\Lambda/3)$ relates to the mass and the distance from the gravitational source. 
Therefore, the parameter $\alpha$ indicates the effect 
of the attractive force of gravity. 
Large $\alpha$ implies that the gravity source is closer or that the gravitational field is stronger. 
Hence, since the conditions of AGPS$_\Lambda$-max with $k=0$ satisfy those of CGPS-max,
an AGPS$_\Lambda$-max with $k=0$ can be regarded as a CGPS-max including the gravitational attraction controlled by the parameter $\alpha$. 
The gravitational attraction lowers the upper limit of the area of the CGPS-max (compare Eq.~\eqref{arealineq3} 
with Eq.~\eqref{arealineq}), which is consistent with the intuitive argument that 
higher energy density reduces the size of cosmological horizon.

\subsection{Attractive gravity probe surface in constant mean curvature slice}
\label{AGPScmc}

Next, we present the definition of {\it the AGPS in a constant mean curvature slice (AGPS$_\Lambda$-cmc)}. 
\begin{definition}
Let us consider the four-dimensional spacetime following the Einstein equation with the positive cosmological constant $\Lambda$. 
Then, we define the {\it attractive gravity probe surface} (AGPS) $S_\alpha$ with a parameter $\alpha$ 
in a spacelike hypersurface $\Sigma$ with $K=3{\sqrt {\Lambda/3}}$ as a compact surface $S_\alpha$ satisfying $k|_{S_\alpha}>0$ and 
\begin{equation}
r^aD_a k|_{S_\alpha} \geq \alpha k^2, \label{AGPSdS2}
\end{equation}
where $\alpha$ is a constant greater than $-1/2$. 
\end{definition}

This definition is motivated by the fact that the Schwarzschild-deSitter spacetime  satisfies Eq. (\ref{CFSratio}). 
Since, in the viewpoint of the spacelike surface $\Sigma$, the definition is the same as the original one \cite{Izumi2021},  
we have the following theorem.
\begin{theorem}
Let us consider the four-dimensional spacetime following the Einstein equation with the positive cosmological constant $\Lambda$ and 
satisfying the dominant energy condition. Then, for the area $A_\alpha$ of the AGPS$_\Lambda$-cmc $S_\alpha$, 
\begin{equation}
A_\alpha \leq 4\pi \left( \frac{3+4\alpha}{1+2\alpha}E_f(S_\alpha) \right)^2 \label{PIAGPS2}
\end{equation}
holds, where $E_f(S_\alpha)$ is defined by Eq. (\ref{Ef}). 
Equality holds if and only if the lapse function for the outward normal direction to $S_\alpha$ is constant on $S_\alpha$, $S_\alpha$ is totally umbilic,
$^{(3)}R$ vanishes,
and the equality in \eqref{AGPSdS2} holds. 
\end{theorem}

The apparent expression is the same with the inequality obtained for asymptotically flat spacetimes \cite{Izumi2021, Izumi2023}. 
The proof is almost similar with the discussion in subSec.~\ref{SecCMCS}. 
Taking the surface integration over $S_\alpha$ for Eq. (\ref{rDk}), the condition (\ref{AGPSdS2}) with Eq. (\ref{3Riccige0}) implies 
\begin{equation}
\left( \alpha +\frac{3}{4} \right) \int_{S_\alpha}k^2dA \leq \frac{1}{2}\int_{S_\alpha}{}^{(2)}RdA=4\pi. \label{integralkDk2}
\end{equation}
Thus, the topology of $S_\alpha$ is $S^2$ and, using the quasi-local mass defined by Eq. (\ref{Ef}), we have  
\begin{equation}
E_f(S_\alpha) \geq \frac{1+2\alpha}{3+4\alpha}\left( \frac{A_\alpha}{4\pi}\right)^{1/2} \geq 0.
\label{Efalpha}
\end{equation}
The minor arrangement gives us Eq. (\ref{PIAGPS2}). 

If one can take the global inverse mean curvature flow 
\footnote{By careful setting with certain assumptions, one can relax the assumption of the inverse mean curvature flow \cite{Izumi2023}. }, where 
the foliation is described by $y$-constant surfaces $\lbrace S_y \rbrace$,
for $\Sigma$, we can show that $E_f(S_y)$ is a monotonically increasing function of $y$ . 
Then, in asymptotically flat spaces, one can show that $E_f(S_\infty)$ becomes the ADM mass \footnote{An energy/mass associated with the 
Killing vector of the background deSitter spacetime has been proposed in Ref. \cite{AD1982}
(one often calls the Abbott-Deser(AD) energy). In general, however, it is turned out that 
the AD energy is composed of the ADM energy part and momentum \cite{Nakao1994}. In any case, there is no consensus 
on the definition of the energy in asymptotically deSitter spacetimes. Nevertheless, one can discuss the ADM mass as long as 
one focuses on asymptotically flat slices.},
that is, 
\begin{equation}
A_\alpha \leq 4\pi \left( \frac{3+4\alpha}{1+2\alpha}m_{\rm ADM} \right)^2.
\end{equation}

We may rewrite Eq.~\eqref{Ef} as 
\begin{equation}
2\int_{S} {}^{(2)}RdA -\frac{64 \pi^{3/2}}{A^{1/2}(S)}E_f(S) = \int_S k^2 dA. 
\label{redefGM}
\end{equation}
One expects $\theta_- \le 0$ inside the cosmological horizon, and then 
Eq. (\ref{expansion-}) gives us $\theta_- \leq 0$ for $k \geq 2H = \sqrt{4\Lambda/3}$. 
In such regions, Eqs. (\ref{integralkDk2}) and \eqref{redefGM} 
give the following theorem. 
\begin{theorem}
Let us consider the four-dimensional spacetime following the Einstein equation with the positive cosmological constant $\Lambda$ and 
satisfying the dominant energy condition. Then, for the area $A_\alpha$ of the AGPS$_\Lambda$-cmc $S_\alpha$ with a parameter $\alpha$ in $\Sigma$ 
satisfying $k|_{S_\alpha} \geq \sqrt{4\Lambda/3}$,  
\begin{equation}
A_\alpha \leq \frac{12\pi}{3+4\alpha}\frac{1}{\Lambda} \label{arealineq4}
\end{equation}
holds. 
Equality holds if and only if, on $S_\alpha$, the first inequality in Eq.~\eqref{Efalpha} 
becomes the equality and $k=\sqrt{4\Lambda/3}$ holds. 
\end{theorem}

Here, the decrease of the upper limit of the area can be seen again, because of the attractive force of gravity.

\section{Summary and discussion}
\label{summary and discussion}

In this paper, we proposed the two types of the cosmological gravity probe surface (CGPS) as an indicator of the gravity in 
spacetimes with positive cosmological constant $\Lambda$ and then show the areal inequality. The definition of the CGPS
 tacitly depends on the time slice and then we had the two formulations, that is, the maximal-slice-based one (CGPS-max) and 
the constant-mean-curvature-based one (CGPS-cmc). Then, we could show the areal inequality $A \leq 12\pi/\Lambda$ on the 
slice with a certain condition for the Ricci scalar. 
The CGPSs correspond to the cosmological horizon in the Schwarzschild-deSitter spacetime. 
Although, in general, the CGPSs do not exactly coincide with the cosmological horizon, 
in most of situations the cosmological horizon can be approximated by the CGPSs.
In Refs. \cite{Boucher1984, Maeda1998} which showed the same inequality for the cosmological horizon, 
the monotonicity of the quasi-local mass or so was used and special foliations was taken. 
In this paper, we would emphasize that we did not take any foliations, but we put the condition specifying 
the CGPSs in the definition, which guarantees the nonnegativity of a quasi-local mass. 
That is to say, our analysis is quasi-local and the structure inside the surface is irrelevant.
The argument in Ref. \cite{Gibbons1999} based on the second variation of the area of the cosmological horizon 
is merely related to ours. 

We also discussed the AGPSs, and show the areal inequality using the quasi-local mass. As the CGPSs, we propose 
the two types of the AGPS, that is, the maximal-slice-based one (AGPS$_\Lambda$-max) and 
the constant-mean-curvature-based one (AGPS$_\Lambda$-cmc). 
In the discussion of AGPS$_\Lambda$s, the distance from the gravitational sources or the strength of 
the gravitational field is represented by a parameter $\alpha$. 
Since an AGPS$_\Lambda$-max with $k=0$ satisfies the conditions for a CGPS-max, it is regarded as a CGPS-max. 
Because the size of the cosmological horizon is finite in the spacetime with a positive cosmological constant, 
we expect that the effect of the gravitational attraction, that is, a dependence on $\alpha$, 
can be seen in the areal inequality for the cosmological horizon, that is, an AGPS$_\Lambda$-max with $k=0$. 
This expectation is correct, that is, 
we can see in Eq.~\eqref{arealineq3} that the upper bound of the area 
of the AGPS$_\Lambda$-max with $k=0$ decreases with $\alpha$ increasing because of the gravitational attraction 
and the case with $\alpha \to -1/2$, which is the limit where we ignore the gravitational attraction, 
corresponds to that of CGPS-max.
Similar contribution can be seen in 
the areal inequality~\eqref{arealineq4} 
for the AGPS$_\Lambda$-cmc.

Meantime, we realized that the existence of the AGPS in asymptotically flat {\it space} 
near the infinity shows us the positivity of the ADM mass. 
It may be possible to relax the definition of AGPS from the original local form to quasi-local 
one (see Eq.~\eqref{QLAGPS}), that is to say, the renewed AGPS. 
Then the positive mass theorem with a certain expression of the ADM mass shows us the existence of 
the renewed AGPS near the infinity (See Appendix A for the details). 

The theorem presented here may tell us that, if the inequality, for instance  Eq.~(\ref{arealineq}), does not hold among observed quantities, 
one of the assumptions does not hold in our Universe. For example, it may indicate that the cosmological constant is not constant and/or 
${}^{(3)}R$ is less than $2\Lambda$. Singularity may exist somewhere, but we need the global analysis to have definite statement.

\ack

T.~S.~and K.~I.~are supported by Grant-Aid for Scientific Research from the Ministry of Education, 
Science, Sports and Culture of Japan (JP21H05182, JP21H05189). T.~S.~is also supported by JSPS 
Grants-in-Aid for Scientific Research (C) (JP21K03551). K.~I.~is also supported by JSPS Grants-in-Aid 
for Scientific Research (B) (JP20H01902) and JSPS Bilateral Joint Research Projects (JSPS-DST collaboration) 
(JPJSBP120227705).

\appendix

\section{AGPS and positivity of ADM mass}
\label{App}

In Ref. \cite{Izumi2021}, the AGPS is defined as a compact 2-surface $S_0$ satisfying $k>0$ and 
\begin{eqnarray}
r^aD_ak \geq \alpha k^2, \label{AGPS0}
\end{eqnarray} 
where $\alpha > -1/2$. 

As one of the expressions for the ADM mass \footnote{See Ref. \cite{Ashtekar1984} for the first equality. 
In the second line, we used the evolution equation for $k$ in $\Sigma$. Note that, using the geometrical identity 
(\ref{rDk}), the ADM mass is written by the Geroch mass as shown in Ref. \cite{Ashtekar1984}, that is, 
$m_{\rm ADM}=\lim_{r \to \infty} \frac{A_r^{1/2}}{64\pi^{3/2}}\int_{S_r}(2{}^{(2)}R-k^2)dA=\lim_{r \to \infty} E(S_r)$.} , we have 
\begin{eqnarray}
m_{\rm ADM} & = & -\lim_{r \to \infty} \frac{A_r^{1/2}}{16\pi^{3/2}}\int_{S_r}{}^{(3)}R_{ab}r^ar^b dA \nonumber \\ 
 & = & \lim_{r \to \infty} \frac{A_r^{1/2}}{16\pi^{3/2}}\int_{S_r}\left[ r^aD_ak+\frac{1}{2}k^2+\tilde k_{ab}\tilde k^{ab}
+\varphi^{-2}({\cal D} \varphi)^2 \right]dA \nonumber \\
& = & 
\lim_{r \to \infty} \frac{A_r^{1/2}}{16\pi^{3/2}}\int_{S_r}\left(r^aD_ak+\frac{1}{2}k^2
\right)dA, 
\end{eqnarray} 
where $r$ corresponds to the radial coordinate near the spatial infinity, $S_r$ is the $r=$ constant surface, 
$A_r$ is the area of $S_r$, ${}^{(3)}R_{ab}$ is the three-dimensional Ricci tensor of $\Sigma$ and, 
in the last equality, we used the fact that $\tilde k_{ab}=O(1/r^2)$ and ${\cal D}_a \varphi=O(1/r^2)$ near the infinity. 

If there is a sequence of the AGPSs $\lbrace S_r \rbrace$ near the spatial infinity, 
one can show the positivity of the ADM mass. On each $S_r$, $r^aD_a k \geq \alpha_r k^2$ and $k>0$ hold, 
where $\alpha_r >-1/2$ is satisfied and $\alpha_r$ approaches $-1/2$ in the limit $r\to \infty$. 
Then, the condition (\ref{AGPSdS2}) tells us that the ADM mass is non-negative as 
\begin{eqnarray}
m_{\rm ADM} \geq \lim_{r \to \infty} \frac{A_r^{1/2}}{16\pi^{3/2}}\int_{S_r}\left(\alpha_r+\frac{1}{2}\right)k^2 dA \geq 0.
\end{eqnarray} 
This is expected from the consideration on the Schwarzschild spacetime in the original work \cite{Izumi2021}. 

Here note that one can relax the condition (\ref{AGPS0}) in the definition of the AGPS as in the following surface integral form
\begin{eqnarray}
\int_{S_\alpha}r^aD_akdA \geq \int_{S_\alpha} \alpha k^2dA. 
\label{QLAGPS}
\end{eqnarray} 
Since the positivity of the ADM mass has been shown in asymptotically flat space \cite{SY1981, Witten1981}, one can see that 
this renewed AGPS exists near the spatial infinity at least.


\end{document}